\documentclass[a4paper,11pt]{article}
\usepackage{pos}

\newcommand{\muR}{\mu_{\rm R}}
\newcommand{\muF}{\mu_{\rm F}}

\title{NNLO QCD predictions for $t\bar{t}W$ production at the LHC}

\author*[a]{Xiang Chen}

\affiliation[a]{Physik-Institut, Universit\"at Z\"urich, \\
  Winterthurerstrasse 190, 8057 Z\"urich, Switzerland}

\emailAdd{xiang.chen@physik.uzh.ch}

\abstract{The production of a top–antitop quark pair in association with a $W$ boson ($t\bar{t}W$) is one of the heaviest signatures currently explored at the Large Hadron Collider (LHC) and the corresponding rates have been found to be consistently higher than the Standard Model predictions, highlighting the need for more accurate theoretical predictions. In this contribution, I present next-to-next-to-leading order (NNLO) predictions for this process, in which, for the first time, the necessary two-loop amplitudes are explicitly evaluated in the generalised leading-colour limit.}

\FullConference{Loops and Legs in Quantum Field Theory (LL2026)\\
12-17, April, 2026\\
Bayreuth, Germany\\}

\begin{document}
\maketitle

\section{Introduction}

The signature $pp \to t\bar{t}W$ stands out among LHC processes for being one of the heaviest that can be studied with current data. The top quark decays almost immediately ($t \to Wb$), so the final state contains a pair of $b$-jets together with three $W$ bosons, translating into a rich multi-lepton topology once the $W$'s decay leptonically. This feature places $t\bar{t}W$ among the key backgrounds for searches targeting new physics with multiple leptons, and makes it one of the rare SM channels that yield same-sign dilepton events~\cite{ATLAS:2018alq,ATLAS:2019fag,CMS:2020cpy}. At the parton level, the process proceeds at leading order via quark--antiquark annihilation of different flavours; consequently, the $W$ is radiated exclusively from the initial-state light quarks, giving the cross section particular sensitivity to the quark PDFs. The process also receives contributions from both QCD and EW higher orders, so that multiple mass scales play a role simultaneously.

Experimental results have been collected by ATLAS and CMS at $\sqrt{s}=8$\,TeV~\cite{ATLAS:2015qtq,CMS:2015uvn} and $13$\,TeV~\cite{ATLAS:2016wgc,CMS:2017ugv,ATLAS:2019fwo,ATLAS:2023gon,ATLAS:2024moy,CMS:2022tkv,CMS:2025iwa}. A recurring feature is that the measured inclusive rates lie above the SM expectation, with indirect hints pointing in the same direction from $t\bar{t}H$~\cite{ATLAS:2019nvo,CMS:2020mpn} and $t\bar{t}t\bar{t}$~\cite{CMS:2023ftu,ATLAS:2023ajo} measurements. This persistent tension calls for increasingly accurate theory.

On the theory front, much progress has been made over the past decade. On-shell NLO QCD predictions became available first~\cite{Badger:2010mg,Campbell:2012dh,Maltoni:2015ena}, followed by the inclusion of EW corrections~\cite{Frixione:2015zaa,Frederix:2017wme}. Soft-gluon effects have been pushed to NNLL accuracy~\cite{Li:2014ula,Broggio:2016zgg,Kulesza:2018tqz,Broggio:2019ewu}. The full off-shell framework, which retains decay kinematics, was developed at NLO initially in QCD~\cite{Bevilacqua:2020pzy,Denner:2020hgg,Bevilacqua:2020srb} and subsequently extended to the combination of QCD and EW effects~\cite{Denner:2021hqi}. Multi-jet merging techniques have also been applied to these off-shell calculations~\cite{Frederix:2012ps,Frederix:2021agh}, and it is against these merged NLO QCD+EW results that most experimental analyses currently compare their data.

The path to NNLO accuracy for $t\bar{t}W$ was opened by Ref.~\cite{Buonocore:2023ljm}, where the first NNLO QCD computation of the inclusive cross section was carried out. The main difficulty in any such calculation is the two-loop virtual amplitude, and the strategy followed in that work was to circumvent a direct computation by resorting to two complementary kinematic approximations: one in which the $W$ boson is taken to be soft (SA), and another in which the top quark is treated as light relative to all remaining scales, i.e.\ a high-energy or massification limit (MA). Conservative error assessments indicated that the uncertainties stemming from these approximations were smaller than the residual scale dependence, yet verifying this conclusion with an explicit two-loop amplitude evaluation has remained an important open question.

In this contribution, based on the results reported in Ref.~\cite{Becchetti:2026ttW}, I describe an NNLO QCD calculation where the two-loop amplitude for $t\bar{t}W$ is computed directly, for the first time, in the generalised leading-colour approximation (LCA). The LCA is fundamentally different from SA and MA: it is a parametric expansion governed by $1/N_c$, and does not invoke any kinematic extrapolation. The $2\to3$ scattering configuration, with massive top quarks in both the initial and final loop states, generates an extraordinary degree of algebraic complexity together with analytic structures --- elliptic curves and nested square roots --- that place this computation among the most demanding applications of state-of-the-art multi-loop technology. The NNLO cross sections obtained with the LCA are found to agree with those of Ref.~\cite{Buonocore:2023ljm}, thereby providing an independent validation of the earlier approximate approach.

\section{Calculation}

\subsection{Organisation of the NNLO computation}

The computational framework is built around the \textsc{Matrix} code~\cite{Grazzini:2017mhc}, which had been extended to $t\bar{t}W$ production in Ref.~\cite{Buonocore:2023ljm} following its successful application to several related channels~\cite{Catani:2019iny,Catani:2019hip,Catani:2020kkl,Buonocore:2022pqq,Devoto:2024nhl}. IR singularities at NNLO are handled through $q_T$-subtraction~\cite{Catani:2007vq}, whose generalisation to processes involving heavy quarks was developed in Refs.~\cite{Bonciani:2015sha,Catani:2019iny,Catani:2019hip}. The real-emission contribution $t\bar{t}W+\text{jet}$ is treated at NLO with dipole subtraction~\cite{Catani:1996jh,Catani:1996vz,Catani:2002hc}, while soft-parton terms specific to massive quarks are provided by the \textsc{Shark} library~\cite{Catani:2023tby,Devoto:2025eyc}. Tree-level and one-loop matrix elements needed throughout the calculation are supplied by \textsc{OpenLoops}~\cite{Cascioli:2011va,Buccioni:2017yxi,Buccioni:2019sur}.

Within the $q_T$-subtraction formalism, the double-virtual contribution enters the NNLO cross section via the hard coefficient $H^{(2)}$. This quantity is obtained by interefering the two-loop amplitude with the Born amplitude, carrying out UV renormalisation and subtracting the predicted IR singularities~\cite{Ferroglia:2009ii}, and finally dividing by the squared Born matrix element. The one-loop amplitudes and the differential equations for the necessary master integrals were already worked out in Refs.~\cite{Becchetti:2025osw,Becchetti:2025qlu}. Hence $H^{(2)}$ constituted the sole missing ingredient, and its computation represents the central new result of this work.

\subsection{Two-loop amplitude}

The partonic channel under study is $u\,\bar{d} \to t\,\bar{t}\,W^+$, with the $W^-$ case obtained by charge conjugation. Diagrams are produced with \texttt{QGRAF}~\cite{Nogueira:1991ex}, and the insertion of Feynman rules together with the colour-algebra decomposition is carried out in \texttt{FORM}~\cite{Ruijl:2017dtg,Davies:2026cci}. The set of diagrams is organised into 19 integral families, three of which are genuine two-loop families once crossing symmetries of the external legs are accounted for~\cite{Becchetti:2025qlu}; the rest reduce to products of one-loop integrals.

The calculation is performed in the LCA, retaining only the colour structures proportional to $N_c^2$, $N_c n_l$, and $n_l^2$, where $N_c=3$ and $n_l=5$ counts the massless flavours. In line with Ref.~\cite{Becchetti:2025osw}, we project the bare two-loop amplitude onto 24 independent tensor structures using the physical-projector technique~\cite{Peraro:2019cjj,Peraro:2020sfm}, working in the 't Hooft--Veltman scheme~\cite{tHooft:1972tcz}. After projection, each amplitude is expressed as a linear combination of scalar Feynman integrals drawn from the families of Ref.~\cite{Becchetti:2025qlu} and their crossings, with rational-function coefficients.

This calculation faces two significant hurdles. The first concerns the Feynman integrals themselves. They contain elliptic curves and nested square roots~\cite{Becchetti:2025qlu}, features that make a straightforward numerical evaluation over the full $2\to3$ phase space impractical. The strategy we adopt, following the methodology of Refs.~\cite{Badger:2024dxo,Badger:2025ljy}, is to expand the integral bases around $\epsilon=0$ and express the coefficients through an (over-)complete set of special functions that obey a system of first-order algebraic partial differential equations. Numerical values for these special functions are obtained by solving the differential equations with \textsc{AMFlow}'s DE solver~\cite{Liu:2022chg}, using high-precision boundary conditions in the physical region also determined with \textsc{AMFlow}~\cite{Liu:2017jxz}. An important advantage of this representation is that UV and IR poles can be subtracted analytically, after which the finite remainder takes the compact form of a linear combination of the special functions multiplied by rational coefficients. To make sure that the pole-subtraction ingredients all refer to the same basis, we recomputed the one-loop LCA amplitude and the mass-counterterm insertion in this framework and validated them against Ref.~\cite{Becchetti:2025osw}. For a typical phase-space point, evaluating the full set of special functions takes between 20 and 120 CPU-minutes.

The second bottleneck is the size of the rational coefficients. To manage it, we perform all rational operations numerically over finite fields~\cite{vonManteuffel:2014ixa,Peraro:2016wsq}, embedding the workflow in a \textsc{FiniteFlow} dataflow graph~\cite{Peraro:2019svx}. The required integration-by-parts reductions~\cite{Tkachov:1981wb,Chetyrkin:1981qh,Laporta:2000dsw} are generated with \textsc{NeatIBP}~\cite{Wu:2023upw}, and the Laurent expansion in $\epsilon$ is performed within the same dataflow setup. We do not attempt to reconstruct the analytic expressions of the rational coefficients; instead, we evaluate them point by point over finite fields and lift the results to rational numbers through the Chinese remainder theorem~\cite{Peraro:2019okx}. Up to 400 finite-field primes are needed per phase-space point, depending on the kinematic configuration.

\subsection{Numerical cross-check and interpolation}

A fully independent numerical validation has been implemented. In this alternative setup, the unpolarised interference with the tree-level amplitude is computed in the conventional dimensional-regularisation scheme. Integral reduction relies on \textsc{Blade}~\cite{Guan:2024byi}, which builds a block-triangular IBP system~\cite{Guan:2019bcx}. The distinguishing feature of this cross-check is that the system is solved numerically with floating-point arithmetic at each point, thus entirely avoiding the finite-field reconstruction that the main approach requires. The basis integrals feeding this system are computed by solving the differential equations of Ref.~\cite{Becchetti:2025qlu} with \textsc{AMFlow}'s DE solver~\cite{Liu:2022chg}. Individual blocks contain $\mathcal{O}(100)$ equations, so numerical error accumulation is not an issue. Additionally, the dimensional regulator $\epsilon$ is treated numerically: we fix $\epsilon = \pm \bar\epsilon$, where $\bar\epsilon = 1/1000$, and keep this numerical value throughout the entire calculation~\cite{Bi:2023bnq,Bi:2025oga,Li:2025bsq}. The finite remainder is then recovered by combining the two evaluations, with a residual uncertainty of $\mathcal{O}(\bar\epsilon^2)$. The two computations agree to five significant digits at all benchmark points tested.

The two-loop finite remainder is sampled on a five-dimensional phase-space grid of $224{,}640$ points, parameterised as suggested in Ref.~\cite{Agarwal:2024jyq} through two energy fractions and three angular variables. A few grid points suffer from strong numerical cancellations and are recomputed with elevated precision to guarantee at least five accurate digits throughout. The hard-virtual coefficient $H^{(2)}$ is then interpolated with quadratic B-splines. The one-loop validation of the interpolation yields sub-permille accuracy for the inclusive integral and percent-level accuracy for differential distributions (referenced to the full NLO cross section). At two loops, the integrated result is essentially insensitive to the precise interpolation scheme within the numerical integration errors.

\section{Phenomenological results}

\subsection{Setup and comparison of approximations}

Predictions refer to proton--proton collisions at $\sqrt{s}=13$\,TeV, chosen to enable a side-by-side comparison with the results of Ref.~\cite{Buonocore:2023ljm}. The top-quark pole mass is fixed at $m_t = 173.2$\,GeV, the $W$-boson mass at $m_W = 80.385$\,GeV, and EW inputs follow the $G_\mu$-scheme with a diagonal CKM matrix. We employ the \texttt{NNPDF31\_nnlo\_as\_0118\_luxqed} PDF set~\cite{Bertone:2017bme}. Central renormalisation and factorisation scales are $\muR=\muF=m_t+m_W/2$, and scale uncertainties are determined from the customary seven-point variation.

\begin{figure}[t]
  \centering
  \includegraphics[width=0.6\textwidth]{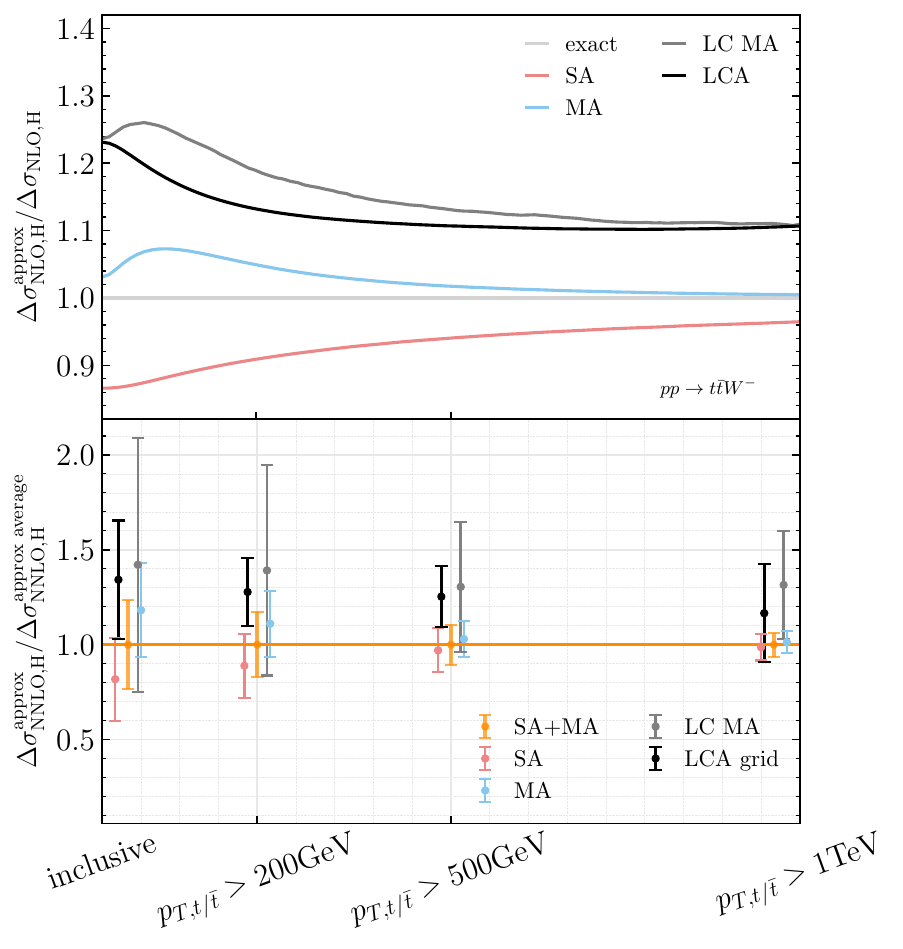}
\caption{
Results for $\Delta\sigma_{\rm NLO,H}$ (upper panel) and $\Delta\sigma_{\rm NNLO,H}$ (lower panel) in the case of $t\bar{t}W^-$ production, for the soft approximation (SA), massification approach (MA) and LCA, as functions of the lower cut applied to the transverse momenta of the top quarks. At NLO the approximations are normalised to the exact result, while at NNLO to the best result (SA$+$MA) from Ref.~\cite{Buonocore:2023ljm}. The uncertainties assigned to each approximation at NNLO are discussed in the text. Similar results are obtained for $t\bar{t}W^+$.
}
  \label{fig:comp}
\end{figure}

Figure~\ref{fig:comp} collects the NLO and NNLO hard-virtual contributions obtained with the LCA, plotted together with the SA and MA predictions of Ref.~\cite{Buonocore:2023ljm} as functions of a lower cut on $p_{T,t/{\bar t}}$. The SA corresponds to the limit of a soft $W$ boson, whereas the MA exploits the high-energy regime where the top mass is negligible relative to all other scales, constructed via massification~\cite{Mitov:2006xs,Wang:2023qbf}. Only the hard-virtual coefficient $H^{(n)}$ receives the approximation; the Born piece stays exact in all cases.

Looking first at the upper panel, which presents NLO predictions relative to the exact one-loop calculation, one observes the pattern established in Ref.~\cite{Buonocore:2023ljm}: SA falls below the exact curve, MA rises above it, and both converge toward the exact result when a hard cut is imposed on the top transverse momenta. In that large-$p_{T,t/{\bar t}}$ domain the MA reproduces the exact result at the percent level. The LCA behaviour is somewhat different --- it overestimates the exact prediction by around $22\%$ at the inclusive level, with this offset shrinking to roughly $10\%$ at large $p_{T,t/{\bar t}}$. For large transverse momenta the LCA is in excellent agreement with its high-energy counterpart, the LC MA, recapitulating the trend observed for the full-colour case.

The lower panel turns to the NNLO comparison, where all curves are divided by the SA$+$MA central value from Ref.~\cite{Buonocore:2023ljm} (orange band). The SA$+$MA result was defined there as the arithmetic mean of the SA and MA predictions, with uncertainties combined linearly. The uncertainty assigned to each of these approximations was derived from its NLO performance: the relative uncertainty at NNLO was taken as twice the relative difference between the approximate and exact NLO results, a conservative rule motivated by the fact that SA and MA are dynamical approximations applied far outside the kinematic domains where they are formally justified. Variations of the IR subtraction scale $\mu_{\mathrm{IR}}$ around $Q$ are also folded in.

\subsection{Uncertainty estimate for the LCA}

The situation is different for the LCA. Being a parametric expansion in $1/N_c$, it does not share the kinematic limitations of SA and MA; its main deficiency is expected to appear near the production threshold, where subleading-colour (SLC) effects become prominent. This kinematic region, however, makes a negligible contribution to the total cross section. Two empirical observations further support a reliable uncertainty estimate: at one loop, the difference between the exact and LCA coefficients shows negligible variation across phase space beyond the threshold region, and the ratio of the two-loop to one-loop LCA coefficients is similarly flat. These facts suggest that the relative one-loop difference provides a fair estimate of the missing SLC effects at two loops.

We therefore estimate the LCA uncertainty on $H^{(2)}$ from the relative discrepancy between the exact and LCA one-loop results. Shifting $\mu_{\mathrm{IR}}$ around $Q$ produces only subleading effects. The numerical error introduced by the grid interpolation, which stays below $1\%$ inclusively but grows to around $12\%$ in the high-$p_{T,t/{\bar t}}$ region, is added linearly. The resulting LCA predictions (black markers in Fig.~\ref{fig:comp}) lie systematically above the SA$+$MA reference, but the two uncertainty bands are fully compatible. Grey markers display the LC MA result, built from massless two-loop $W+4$-parton amplitudes~\cite{Abreu:2021asb,Badger:2021nhg} with the massification ingredients kept in LC.

\subsection{Integrated cross sections}

The inclusive numbers for the hard-virtual contribution $H^{(2)}$ are listed in Table~\ref{tab:H2}.

\begin{table}[h]
\centering
\begin{tabular}{lcc}
\hline
& $t\bar{t}W^-$ [fb] & $t\bar{t}W^+$ [fb] \\
\hline
SA    & $15.67\phantom{0}(0)\pm 4.21$ & $34.59\phantom{0}(0)\pm \phantom{0}9.18$ \\
MA    & $22.62(15)\pm 4.78$ & $49.87(40)\pm \phantom{0}8.74$ \\
SA$+$MA & $19.14\phantom{0}(8)\pm 4.49$ & $42.23(20)\pm \phantom{0}8.96$ \\
LCA   & $25.70\phantom{0}(0)\pm 5.68$ & $56.76\phantom{0}(0)\pm 12.03$ \\
\hline
\end{tabular}
\caption{Contribution of $H^{(2)}$ to the inclusive NNLO cross section in the various approximations. The numbers in parentheses indicate the Monte Carlo integration error; the second uncertainty is the estimated uncertainty on each approximation. For SA and MA this estimate follows the prescription of Ref.~\cite{Buonocore:2023ljm}; for the LCA it also includes an estimate of the grid-interpolation error, as discussed in the text.}
\label{tab:H2}
\end{table}

As expected from Fig.~\ref{fig:comp}, the LCA values for $H^{(2)}$ are the largest among all approximations, but still consistent with the others inside the quoted errors. When the two-loop term is combined with the exactly computed remaining NNLO pieces, one arrives at the full NNLO QCD cross sections presented in Table~\ref{tab:xs}.

\begin{table}[h]
\centering
\begin{tabular}{lcc}
\hline
& $t\bar{t}W^-$ [fb] & $t\bar{t}W^+$ [fb] \\
\hline
SA$+$MA~\cite{Buonocore:2023ljm} & $235.4\,(0)^{+5.1\%}_{-6.6\%}\pm 1.9\%$ & $474.9\,(2)^{+4.8\%}_{-6.4\%}\pm 1.9\%$ \\
LCA & $241.9\,(0)^{+6.4\%}_{-7.3\%}\pm 2.3\%$ & $489.4\,(1)^{+6.3\%}_{-7.2\%}\pm 2.5\%$ \\
\hline
\end{tabular}
\caption{NNLO QCD cross sections. The perturbative uncertainties are estimated through standard seven-point scale variation. The additional uncertainty assigned to the SA$+$MA result corresponds to the approximation error estimated in Ref.~\cite{Buonocore:2023ljm}; for the LCA it accounts for the missing subleading-colour contributions and the grid-interpolation error, as discussed in the text.}
\label{tab:xs}
\end{table}

The LCA-based central values exceed the SA$+$MA reference by approximately $3\%$, well within the respective uncertainties. Comparing with Table~\ref{tab:H2}, the two-loop virtual contribution itself is substantial: it represents about $11\%$ of the full NNLO cross section.

\section{Summary}

This contribution has covered an NNLO QCD calculation of $t\bar{t}W$ hadroproduction where, for the first time, the two-loop virtual amplitudes are directly evaluated, in the generalised LCA. The full computation is documented in Ref.~\cite{Becchetti:2026ttW}. Carried out on a phase-space grid of more than $200{,}000$ points using state-of-the-art multi-loop amplitude techniques, the calculation has been validated through a completely independent numerical approach.

The resulting NNLO cross sections fully corroborate the earlier determination of Ref.~\cite{Buonocore:2023ljm}, which had relied on dynamical approximations of the two-loop contribution. The uncertainty associated with the unresolved subleading-colour pieces is estimated at roughly $2.5\%$ --- slightly above the previous estimate, yet still subdominant compared with scale variations. Going beyond the phenomenological impact, this work shows that numerical methods rooted in special-function representations have reached a level of maturity where they can be deployed in complete two-loop calculations for complex LHC final states. In the near term, the LCA results provide a basis for approximating the subleading-colour effects following the roadmap of Ref.~\cite{Buonocore:2023ljm}, for which all the required building blocks are now in place.

\section*{Acknowledgments}
\vspace{-0.2cm}
I would like to thank M. Becchetti, D. Canko, V. Chestnov, M. Delto, S. Ditsch, M. Grazzini, S. Kallweit, T. Peraro, M. Pozzoli, C. Savoini, L. Tancredi, and S. Zoia for the excellent collaboration on which the results presented here are based. This work is supported by the Swiss National Science Foundation (SNSF) under contract 200020\_219367, and by the UZH Postdoc Grant, grant No.~[FK-25-104].

\bibliographystyle{UTPstyle}
\bibliography{bibliography}

\end{document}